\documentclass{osa-article}

\journal{oe}

\begin{document}

\title{Coexisting frequency combs spaced by an octave in a monolithic quantum cascade laser} 

\author{Andres Forrer,\authormark{1} Markus R\"osch,\authormark{1} Matthew Singleton,\authormark{1} Mattias Beck,\authormark{1} J\'er\^ome Faist,\authormark{1} AND Giacomo Scalari\authormark{1,*}}

\address{\authormark{1}ETH Zurich, Institute of Quantum Electronics, Auguste-Piccard-Hof 1, Zurich 8093, Switzerland}

\email{\authormark{*}scalari@phys.ethz.ch} 



\begin{abstract}
Quantum cascade lasers are proving to be instrumental in the development of compact frequency comb sources at mid-infrared and terahertz frequencies. Here we demonstrate a heterogeneous terahertz quantum cascade laser with two active regions spaced exactly by one octave. Both active regions are based on a four-quantum well laser design and they emit a combined 3~mW peak power at 15~K in pulsed mode. The two central frequencies are 2.3~THz (bandwidth 300~GHz) and 4.6~THz (bandwidth 270~GHz). The structure is engineered in a way that allows simultaneous operation of the two active regions in the comb regime, serving as a double comb source as well as a test bench structure for all waveguide internal self-referencing techniques. Narrow RF beatnotes ($\sim$ 15~kHz) are recorded showing the simultaneous operation of the two combs, whose free-running coherence properties are investigated by means of beatnote spectroscopy performed both with an external detector and via self-mixing. Comb operation in a highly dispersive region (4.6~THz) relying only on gain bandwidth engineering shows the potential for broad spectral coverage with compact comb sources.
\end{abstract}


\section{Introduction}

Frequency combs\cite{Udem2002} (FCs) have revolutionized many fields in physics such as metrology, spectroscopy and astronomy \cite{Udem2002,steinmetz2008,diddams2010evolving,coddington2016}. FCs act as rulers in the frequency domain as they consist of a set of perfectly equally spaced phase-locked modes. Especially for molecular sensing there is a strong interest to have compact and powerful FCs at mid-infrared (mid-IR) and terahertz (THz) frequencies where most rotational and vibrational absorption lines are located \cite{tonouchi2007,villares2014,faist2016combs}. FCs generated by quantum cascade lasers (QCLs) are promising sources to fulfill this demand both in the mid-IR \cite{Hugi2012,villares2014,faist2016combs}, as well as in the THz frequency region \cite{Burghoff2014,Roesch2014,faist2016combs}. QCLs are unipolar semiconductor lasers which potentially allow very compact FC setups. First FC spectroscopy experiments using QCLs have already been performed using a so-called dual-comb spectrometer \cite{villares2014,yang2016,APLwestberg_wisocky2017}. It has also been shown that such a dual-comb source can be shrunk down to chip scales \cite{villares2015onchip,Roesch2016}. 
However, an open issue in QCLs is the full comb stabilization. By definition, a FC consists of equidistantly spaced and phase-locked modes. The frequency of each mode $n$ can therefore be written as
\begin{equation}
 f_n=f_{ceo}+n\cdot f_{rep},
 \label{eq:comb}
\end{equation}
with $f_{ceo}$, $f_{rep}$ being the carrier envelope offset frequency and the repetition frequency of the laser cavity respectively. From Eq. \eqref{eq:comb} it is evident that both $f_{ceo}$ and $f_{rep}$ need to be measured and controlled in order to fully stabilize the entire FC. Only a full stabilization of the FC will give access to the maximal frequency accuracy given by the linewidth of the individual comb teeth lines \cite{cappelli2016,faist2016combs}. In QCL FCs, one has direct access to the repetition frequency, which can be extracted on the bias line of the laser \cite{villares2014,Roesch2016}. A stabilization can be achieved by introducing a small change on the driving current according to the measured shift of the repetition frequency using a control loop \cite{villares2014,Burghoff2014}. On the contrary, there is not such a direct access of $f_{ceo}$ over the bias line. The so-called f-2f technique \cite{telle1999f2f,jones2000f2f,diddams2000}, where a FC spanning more than an octave is required, can provide a measurement of $f_{ceo}$ and therefore the possibility to lock it to a microwave reference source. The key idea of this technique is to frequency double a low frequency mode of the FC by a $\chi ^{(2)}$ nonlinear medium and let it beat with the closest mode in the high frequency part of the FC, leading to a RF beating signal at $f_{ceo}$. THz QCLs offer interesting possibilities in the respect of this f-2f technique since they can yield emission bandwidths of more than one octave \cite{Roesch2014,Roesch2017Nanophot}, but not yet phase coherent over the whole spectrum. The challenge is mainly constituted by the dispersion compensation which, in fully integrated semiconductor lasers is far from trivial. The frequency doubling could be implemented internally in a QCL thanks to the possibility of designing high second order nonlinearities, as demonstrated by Gmachl \textit{et~al.} in Ref. \cite{Gmachl:JQE:03:1345}.

In this paper we explore two coexisting FCs spaced by one octave and simultaneously operating in the same waveguide. Only recently, two comb lasing based on heterogeneous waveguide coupling has been reported \cite{yang_lateral_2018}.  Here we employ our 4-well active region \cite{amanti2009} by rescaling and centering the two emission frequencies at 2.3~THz and 4.6~THz, in order to have high intensities and stable phase coherent operation independently at the octave spacing. This way to operate is different from our previous works where we achieved octave-spaced laser lines \cite{Roesch2014,Roesch2017Nanophot}, not in a comb regime, from the low intensity modes on the sides of the emission of a three or four stack heterogeneous laser. 

\section{Results and discussion}
\subsection{Laser design and performance}

\begin{figure}[tb]
  \centering
  \includegraphics[width=0.8\linewidth]{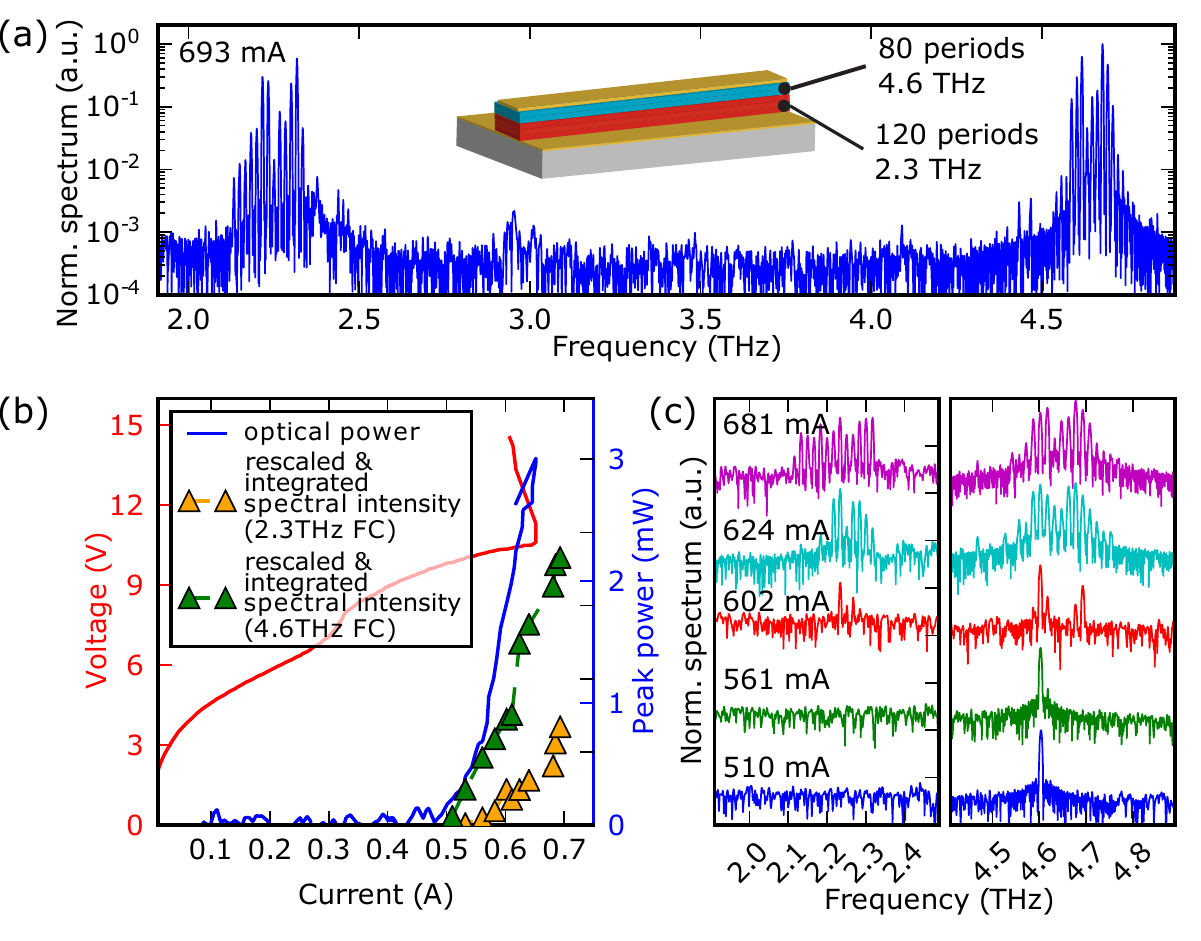}
  \caption{(a) Spectrum showing the simultaneous lasing at octave-spaced frequencies in pulsed operation (5\% duty cycle at 1~kHz repetition frequency at 15~K) and in the inset the schematic of the heterogeneous QCL. (b) L-I-V curve for a 2.4~mm long, 80~$\mu m$ wide laser ridge in pulsed mode (at 120~kHz and 5\% duty cycle, square wave modulated at 30~Hz at 15~K). Peak power of 3mW is recorded. The orange resp. green marker show the spectral integrated intensities from (c) around the 2.3~THz resp. 4.6~THz FC, rescaled to the peak power, giving an estimate for the individual FC peak power in the mW range. (c) Series of recorded spectra as a function of the injected current of the same device operated as in (a). The evolution shows first lasing of the 4.6~THz active regions before both active regions lase simultaneously.}
  \label{fig:1}
\end{figure}

With the heterogeneous QCL design we can study the simultaneous presence of two octave-spaced combs operating in a common waveguide. The alignment electric field values for the two active regions are very different, but this is not a problem for the simultaneous lasing of both stacks: what is relevant is the threshold current density, which has to be very similar for the alignment conditions of both stacks. Both structures were previously studied as homogeneous lasers and the respective layer sequences are reported by K. Otani \textit{et~al.} in Ref. \cite{OtaniNJP201647THz} and J. Lloyd-Hughe \textit{et~al.} in Ref. \cite{Lloyd-Hughes:09}. They are adjusted to operate in the same current density range acting mainly on the injection barrier thickness which is 5.5~nm for the 2.3~THz structure and 5.1~nm 4.6~THz. The double metal waveguide is composed of 120 repetitions of the 2.3~THz active region (sheet density $n_s^{2.3THz}=3.6 \times 10^{10}$ cm$^{-2}$ ) and 80 repetitions of the 4.6~THz active region (sheet density $n_s^{4.6THz}=3.3 \times 10^{10}$ cm$^{-2}$ ), for a total thickness of 13~$\mu$m. A schematic of the QCL is shown in the inset of Fig. \ref{fig:1}(a).

A L-I-V plot for a typical ridge device of 2.4~mm length and 80~$\mu m$ width provided with metal setback and side absorbers (see Bachmann \textit{et~al.} in Ref. \cite{Bachmann:16}) is shown in  Fig. \ref{fig:1}(b) with peak power of 3 mW. The QCL is operated in pulsed mode, at 120~kHz repetition frequency, 5\% duty cycle and square wave modulated by 30~Hz at 15~K (detection with an \textit{Absolute Terahertz Power Meter} by \textit{Thomas Keating Ltd}). The markers represent the integrated spectral intensity of the 2.3~THz (orange) resp. 4.6~THz FC (green), normalized to the L-I curve; here the device was operated at 1~kHz and 5\% duty cycle. This provides an estimate of the power distribution between the two FCs. The difference in the dynamical range between L-I and rescaled integrated spectral intensities curves is due to different pulse duration settings for each kind of measurement. Fig. \ref{fig:1}(c) reports the spectral evolution of the emitted radiation as a function of the injected current and Fig. \ref{fig:1}(a) shows a connected spectrum at 693 mA injection current. Therefore clear laser signals at 2.3~THz and 4.6~THz with bandwidths of the order of 300~GHz are recorded.

\begin{figure}[tb]
	\centering
	\includegraphics[width=0.8\linewidth]{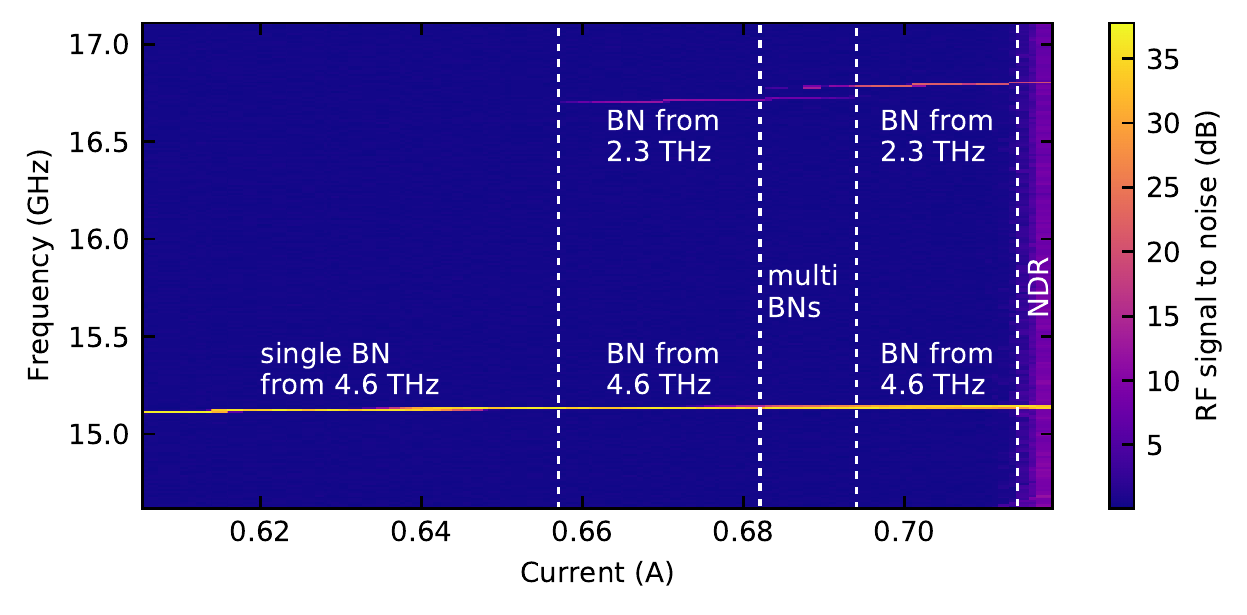}
	\caption{BN map from a 2.4~mm long and 80~$\mu m$ wide QCL operated in pulsed (1~kHz repetition rate, 5 \% duty cycle) at 13.3~K. We observe two regions where two single BNs at 15.1~GHz resp. 16.8~GHz are simultaneously present with a SNR of $\sim$35 dB resp. $\sim$ 20 dB. The assigning of the BNs was done by the simple argument of the expected $f_{rep}$ from the frequency-dependent free spectral range of the QCL and is further addressed in the text.}
	\label{fig:2}
\end{figure}

To have a first indication for FC emission, a beatnote (BN) should be measurable. QCLs offer the possibility to access these BNs directly over the bias line and detection by a RF spectrum analyzer (SA; R\&S, FSW26). By scanning the current and simultaneously recording the RF spectrum a BN map can be generated. Such a BN map is shown in Fig. \ref{fig:2} for a 2.4~mm and 80~$\mu m$ wide QCL in a pulsed configuration (5 \% duty cycle at 1~kHz repetition frequency) at 13.3~K. It shows first a single BN regime at 15.1~GHz followed by two regimes, separate by a short multi BN region, with two clear single BNs spaced by roughly 1.7~GHz. The main interest is the injection current range, where two strong and narrow BNs for the coexisting FCs are present. The assignment of the BNs to the FCs in Fig. \ref{fig:2} is discussed in the next paragraph and verified in section \ref{ComCoherence}. The tested QCLs also emit in CW operation showing only single mode lasing or single FC regimes and were therefore not studied further.

\begin{figure}[tb]
	\centering
	\includegraphics[width=0.8\linewidth]{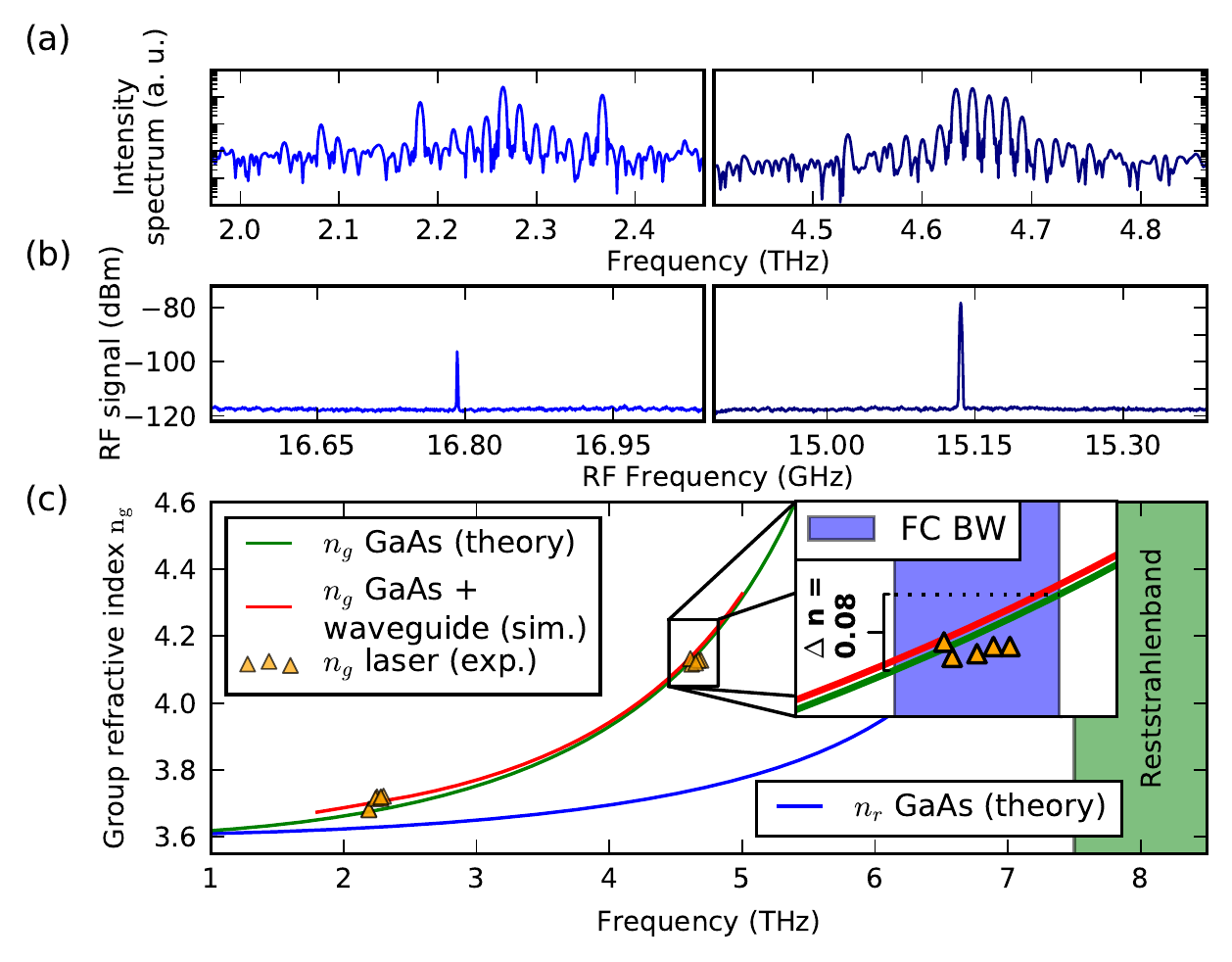}
	\caption{(a) Spectrum of the two simultaneous existing FCs at 710 mA injection current of the BN map in Fig. \ref{fig:2}. (b) Corresponding narrow BNs recorded through the bias-tee. (c) Calculated group refractive index $n_g$ for GaAs (green) and the simulated group refractive index for the QCL waveguide (red, using \textit{Comsol 5.0}) together with values deduced from BNs measurement of the two combs, using $n_g=c/(2 f_{rep} L)$ for multiple devices operating in two color comb regime. Inset: Multiple devices show lasing in a very dispersive region. The recorded FC BW shows the capability of the QCL of establishing a FC despite the approximate 0.08 change in the group refractive index. 
	}
	\label{fig:3}
\end{figure}

At 710 mA injection current, where two BNs are present in Fig. \ref{fig:2}, we measure a close up of the BNs over the bias-tee and record the spectrum of the QCL by a commercial FTIR (Bruker 80v, DTGS detector). The corresponding intensity spectrum is shown in Fig. \ref{fig:3}(a), leaving away inter comb frequencies. Two distinct BNs are observed at frequencies of 16.79~GHz and 15.14~GHz as shown in Fig. \ref{fig:3}(b). Free-running BN linewidths of tenths of kHz are observed on the SA and further analyzed in section \ref{sec:BNdyn}. The difference in the BN frequencies results from the high material dispersion of the GaAs which increases as the laser frequency approaches the reststrahlen band. The simple model of Fabry-P\'{e}rot cavity mode spacings leads to the description of $f_{rep} = c/(2\cdot n_g(\nu)\cdot L)$, where c is the speed of light, L the length of the cavity and $n_g(\nu)$ the frequency dependent group refractive index of the QCL structure.
We investigate systematically, via BN ($f_{rep} \rightarrow n_g(\nu_c)$) and FTIR measurements (FC center frequency $\nu_c$), the mode spacing at 2.3~THz and 4.6~THz for a 2~mm long and 100~$\mu m$ wide and two 2.4~mm long and 80~$\mu m$ wide QCLs, where two simultaneous BNs are present. The assignment of the group index to the FC center frequency is done due to relatively narrowband FCs compared to the FC spacing. We report the results in Fig. \ref{fig:3}(c) together with the calculated material dispersion for GaAs from the book \cite{Palik::85:429} of E. Palik and the simulation for a QCL waveguide (\textit{Comsol 5.0}).

The result presents the potential of QCLs to form FCs in a very dispersive regime, here the 4.6~THz FC near the reststrahlen band of GaAs. As can be seen in the inset of Fig. \ref{fig:3}(c), the bulk group refractive index change over the bandwidth of the FC is estimated to be $\Delta n \approx 0.08$. Only recently QCL FC formation in the dispersive regime at 4.6~THz has been reported by Y. Yang \textit{et~al.} in Ref.\cite{yang_lateral_2018}.

\subsection{Comb coherence properties}
\label{ComCoherence}

\begin{figure}[tb]
	\centering
	\includegraphics[width=0.7\linewidth]{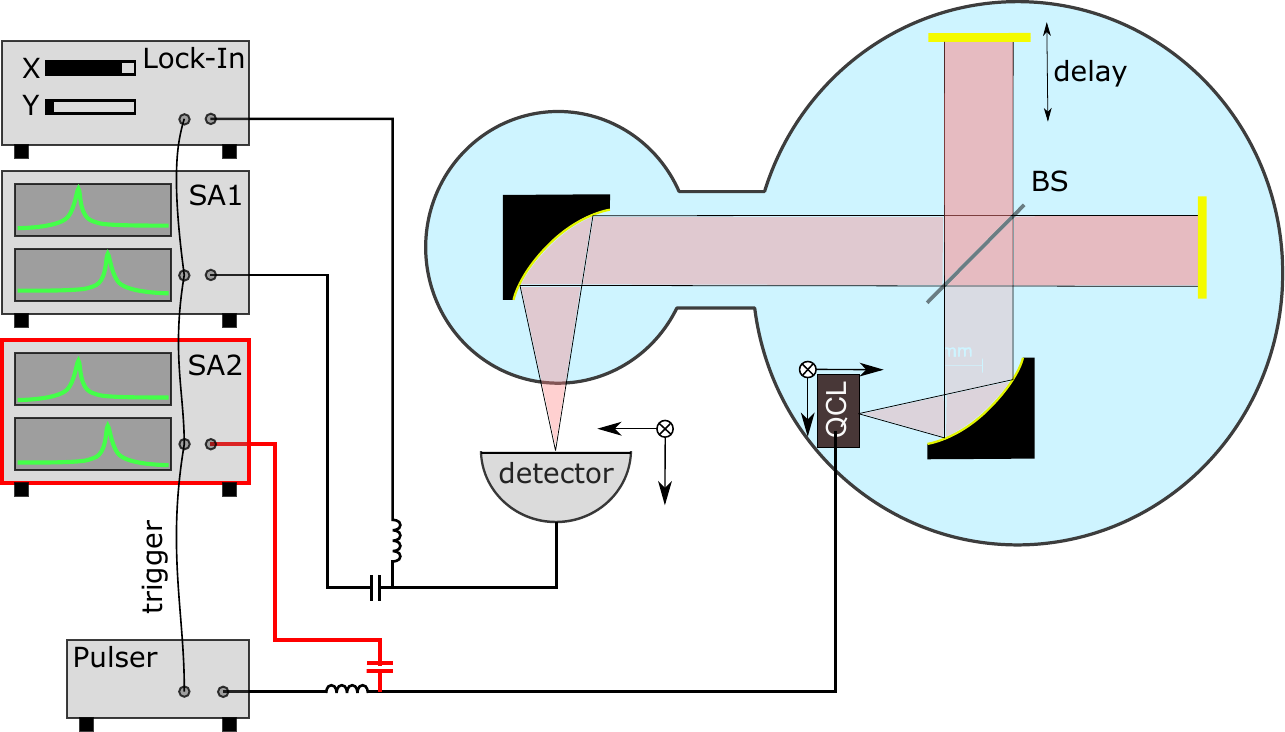}
	\caption{Schematics of intermode BN spectroscopy setup. The SA on the bias line (SA2; highlighted in red) in the sketch can be used to verify the presence of to BNs as well as for detection of the self-mixing signal from the intermode BN spectroscopy. The laser is directly operated on a flow crysotat inside a home-made, step-scan and under vacuum FTIR (light blue area).}
	\label{fig:4}
\end{figure}

The recording of a BN gives incomplete information when assessing and proving the full comb nature of the laser emission. The coherence has to be verified among all the modes which are constituting the laser spectrum because it is not a priori clear which modes contribute to the BN signal. Due to the absence of pulses in the QCL output in the comb regime, standard autocorrelation techniques relying on non-linear processes allowed by high intensities are not an option to show this coherence. To this extent, a few experimental techniques, all based on the concept of spectral filtering combined with RF BN detection have been developed \cite{Hugi2012,faist2016combs,Burghoff2014}. Here we employ the intermode BN spectroscopy technique using a fast THz Schottky detector (\textit{VDI Inc.}) coupled to a home-made step-scan FTIR. The detector allows the simultaneous recording of the RF intermode beating as well as the DC response to the optical THz signal as a function of delay and using appropriate triggering to the pulse. A sketch of the setup is shown in Fig. \ref{fig:4}.

\begin{figure}[tb]
	\centering
	\includegraphics[width=0.7\linewidth]{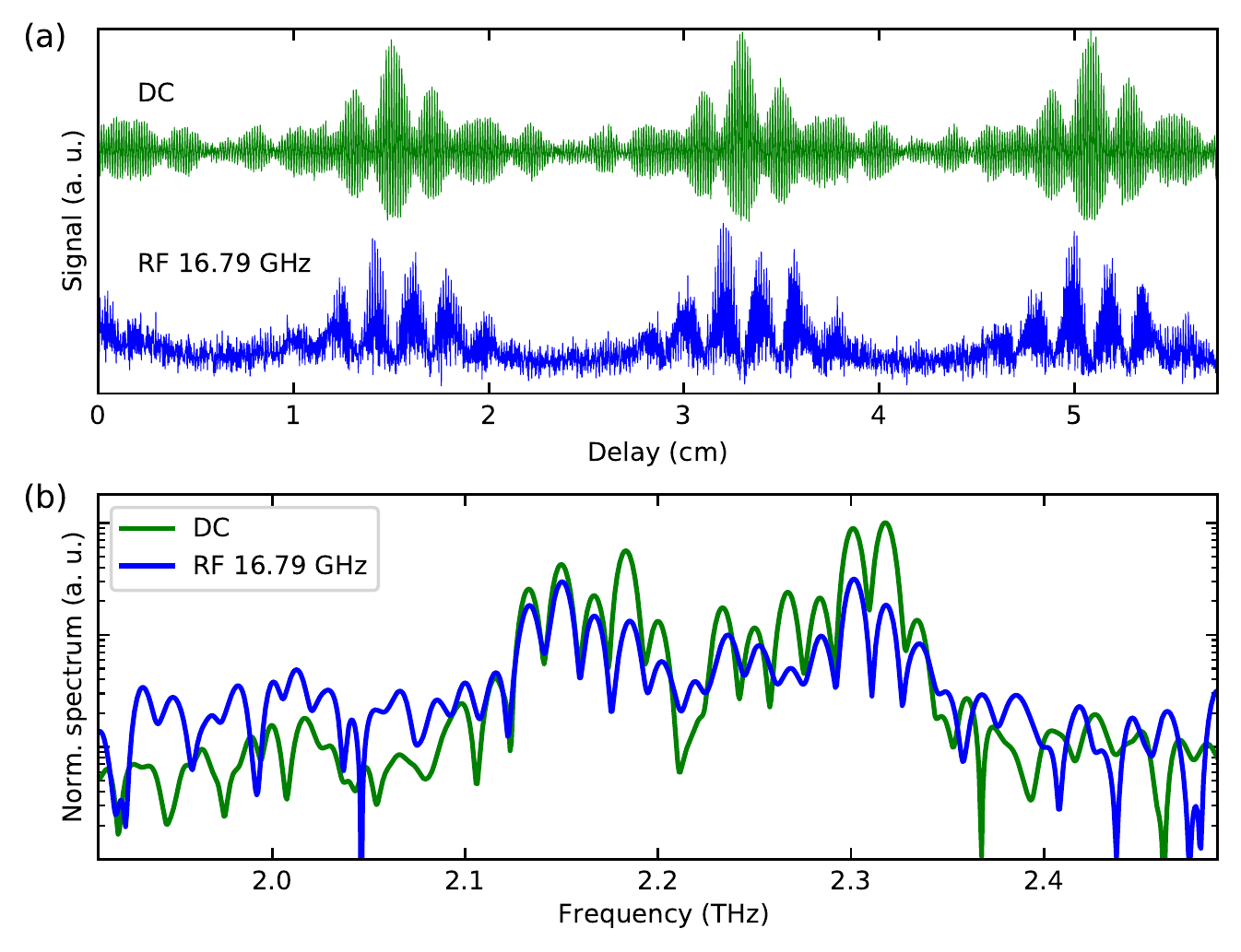}
	\caption{A 2.4~mm and 80~$\mu m$ wide QCL was operated at 1~kHz repetition frequency with 100~$\mu s$ long pulses at 16K. (a) Interferograms recorded from the DC signal (green) and RF signal (blue) of the Schottky detector. (b) Corresponding spectra showing the spectral coherence and confirming the comb nature around 2.3~THz.}
	\label{fig:5}
\end{figure}

A 2.4~mm and 80~$\mu m$ wide QCL was operated in pulsed at 10\% duty cycle at 1~kHz repetition frequency at 16~K. Fig. \ref{fig:5}(a) shows the two simultaneous by the Schottky recorded interferograms of the DC and the RF component at 16.79~GHz of such an measurement. The RF interferogram is deduced from the power change of the intermode beating signal in the RF spectrum for each delay step. The shape of the interferograms qualitatively agree with the mid-IR QCL measurement and model presented by A. Hugi \textit{et~al.} in Ref. \cite{Hugi2012}, indicating the frequency-modulated (FM) operation of the QCL comb. The similarity of the intensity spectra of the two interferograms shown in Fig. \ref{fig:5}(b) confirm the spectral coherence and prove the comb nature for the spectral components centered around 2.3~THz. At the same time it was verified that the second BN at 15.14~GHz was present, indicating lasing of the 4.6~THz FC.

Due to the narrow bandwidth (2.2-3.2 THz) of the antenna, which couples to the Schottky detector, the detection of the radiation around 4.6~THz is highly inefficient and inhibits the intermode BN spectroscopy for the FC centered at 4.6~THz. To assign the BN at 15.14~GHz to the 4.6~THz FC we record the self-mixing signal, i.e. change in BN power due to back-scattered light, as the FTIR is stepping. This approach is similar as the one used by M. Wienold \textit{et~al.} in Ref. \cite{Wienold2014}, where the technique was demonstrated for a single plasmon waveguide THz QCL. In our case we use a double metal waveguide which is inherently less prone to optical feedback with respect to the single plasmon case. 

Fig. \ref{fig:4} shows the setup employed for the self-mixing configuration by using one SA (highlighted in red) for the detection of the self-mixing signal of the BNs. This self-mixing technique together with our intermode BN spectroscopy setup allows us to directly verify the origin of the 15.14~GHz BN from modes centered at 4.6~THz.

\begin{figure}[tb]
  \centering
  \includegraphics[width=0.8\linewidth]{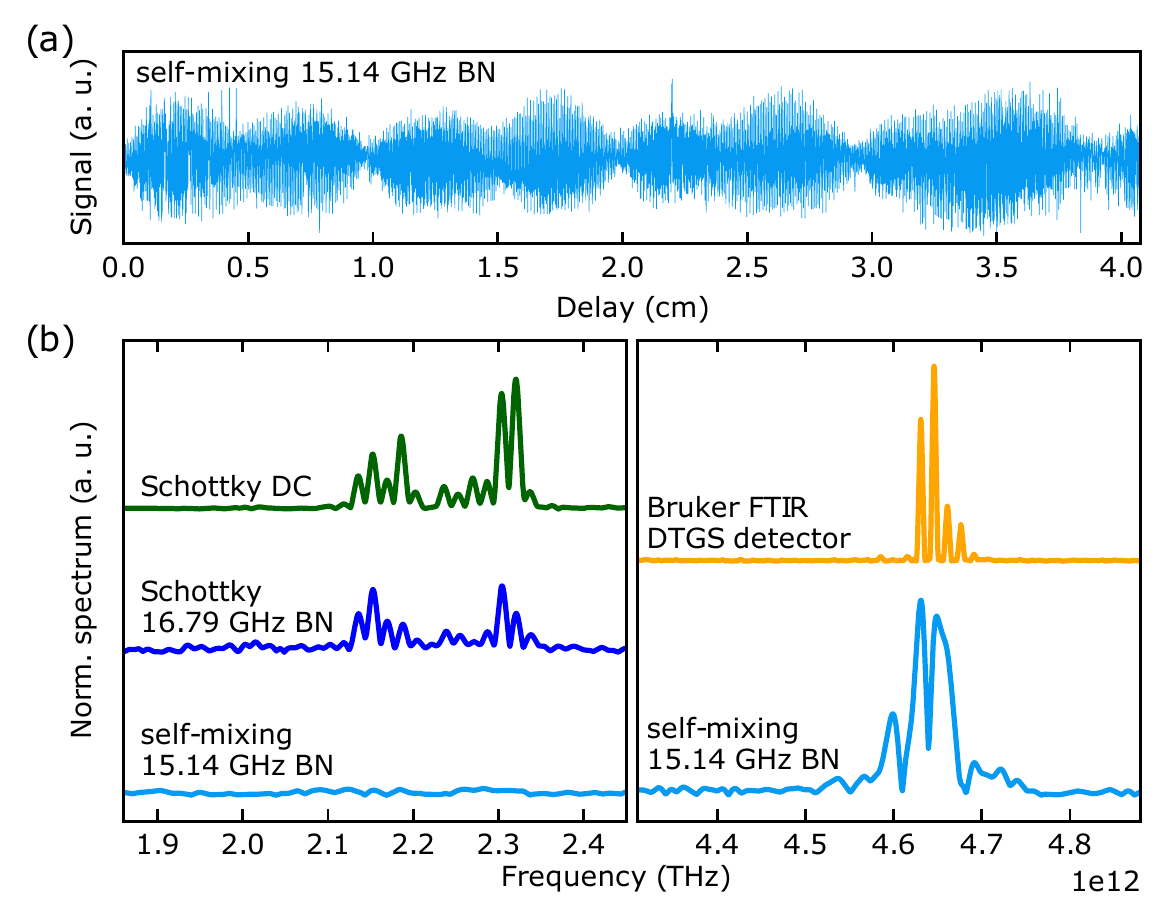}
  \caption{(a) Self-mixing interferogram recorded from the 15.14~GHz BN in a step-scan FTIR. (b) Corresponding spectrum (azure blue) is compared to vertical displaced spectra in the two FC regions. Around 2.3~THz this self-mixing spectrum is clearly not generated by the modes of the 2.3~THz FC as measured by the DC (green) and RF (blue) spectrum from the Schottky measurement. Due the lacking detection bandwidth of the Schottky, the 4.6~THz components of the self-mixing spectrum is compared to a reference measurement by a commercial FTIR (Bruker 80v, DTGS detector). Despite the low resolution of the self-mixing spectra, the origin of the 15.14~GHz BN from modes centered at 4.6~THz is clearly verified.}
  \label{fig:6}
\end{figure}

In Fig. \ref{fig:6}(a) we report a separate self-mixing interfrogram from the 15.14~GHz BN, additionally to the DC and RF interferogram in Fig. \ref{fig:5}(a). The self-mixing method can also be applied for the 2.3~THz FC generating the 16.79 BN, but suffers from lower sensitivity and linearity of the detection compared to the optical RF detection by the Schottky detector and is therefore omitted from the comparison. In Fig. \ref{fig:6}(b) we compare the self-mixing spectrum from the 15.14~GHz BN to the DC and optical 16.79~GHz RF detection and can clearly verify that, as expected, the 2.3~THz FC does not generate the 15.14~GHz BN. Due to the lacking detection bandwidth of the Schottky detector at 4.6~THz, the 15.14~GHz BN self-mixing spectrum is compared to an additional spectrum recorded in a commercial FTIR (Bruker 80v). Despite the lower resolution due to shorter delay range and lower SNR of the self-mixing detection, the two spectra show a qualitative overlap and therefore the origin of the 15.14~GHz BN. In summary we verified the mode coherence in the 2.3~THz comb, its BN generation at 16.79~GHz and the origin of the 15.14~GHz BN generated by the comb located at 4.6~THz.

\newpage
\subsection{Beatnote $\mu s$ dynamics}
\label{sec:BNdyn}

\begin{figure}[tb]
	\centering
	\includegraphics[width=0.8\linewidth]{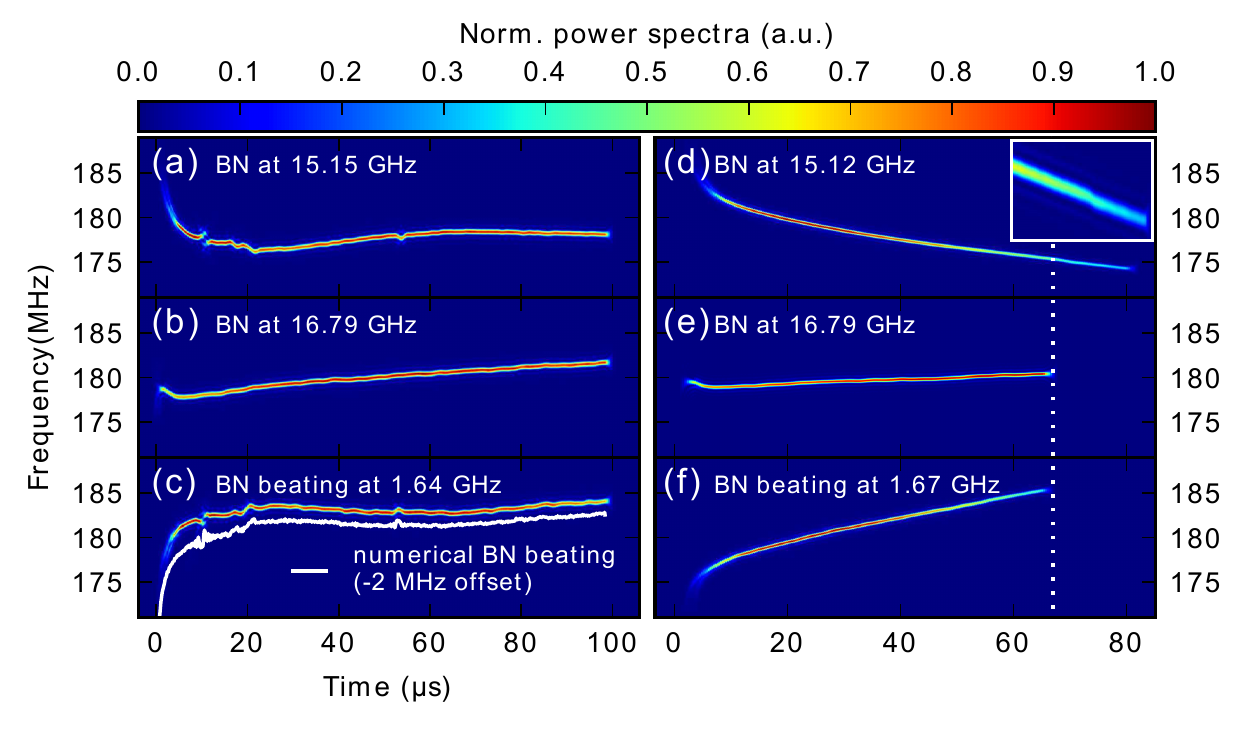}
	\caption{(a) Down-mixed 15.15~GHz BN from a 2.4~mm device at 697~mA injection current operated at 500~Hz repetition frequency and 100~$\mu s$ long pulses at 7~K. The BN shows red and blue shift during the same pulse. (b) Down-mixed signal from the simultaneous existing 16.79~GHz BN showing a blue shift. (c) Down-mixed signal from the BN beating at 1.64~GHz and the numerical BN beating (white) from (b) and (b), offset by -2 MHz for visibility.
	(d) Down-mixed 15.12~GHz BN from the same 2.4~mm device at 668~mA injection current operated at 1~kHz repetition frequency and 80~$\mu s$ long pulses at 7.7~K. The BN shows a red shift. Inset: Slight change of the 15.12~GHz BN trace when the 16.79~GHz BN switches off. (e) Down-mixed signal from the simultaneous existing 16.79~GHz BN showing a blue shift. The BN switches off after $\sim$67~$\mu s$ (dotted white line). The slight change at this time in the time trace of the 15.12~GHz signal could indicate coupling between the BNs. (f) Down-mixed signal from the BN beating at 1.64~GHz. The switching-off of the 16.79~GHz BN is reflected in the BN beating.}
	\label{fig:7}
\end{figure}

In a second part of the study we investigate briefly the long-term dynamics, here $\mu s$ range, of the BNs generated by the pulsed THz comb. The BNs are extracted over a bias-tee in the bias line, down-mixed to 180 MHz on the lower sideband and recorded by a fast oscilloscope (2.5 GS/s). The spectrograms for both down-mixed BNs arising from the 2.3 and 4.6~THz FCs as well as the down-mixed beating signal of the BNs for two different operation points of the same 2.4~mm long device are shown in Fig. \ref{fig:7}. Fig. \ref{fig:7}(a-c) shows the successively recorded down-mixed BNs from the 2.3~THz FC, 4.6~THz FC and their beating, operating at 697~mA injection current at 500~Hz repetition frequency, 100~$\mu s$ long pulses at 7~K. The 15.15~GHz BN from the 4.6~THz FC in Fig. \ref{fig:7}(a) shows red and blue shifts during one pulse. The corresponding 16.79~GHz BN signal of the 2.3~THz FC is represented in Fig. \ref{fig:7}(b), showing a blue shift. The frequency difference of the two BNs at 1.64~GHz is also detected, due to non-linear mixing inside the QCL cavity, most probably rectification. The signal is shown in Fig. \ref{fig:7}(c). In white we present the numerical beating of the successively recorded traces in (a) and (b), offset by -2 MHz for better visibility, verifying the BN beating. The BN drifts show therefore that future CW operation could be favored depending on the $f_{rep}$ stability requirements.

In Fig. \ref{fig:7}(d-f) the same measurement is repeated for a different operation point, namely 668~mA injection current at 1~kHz repetition frequency, 80~$\mu s$ long pulses at 7.7~K.  The 15.12~GHz BN is presented in (d) and shows this time a pure red shift. The 16.79~GHz BN in (e) shows a similar behavior as in (b), but switches off before the pulse ends. Surprisingly we see a slight change in the BN time trace in (d) at this time (see Fig. \ref{fig:7}(d) inset). This could indicate a weak coupling of the BNs and is still under further investigation. Switching-off of one BN as observed and presented in Fig. \ref{fig:7}(e) is a unique feature at this specific operation point and in most of the other investigated time-traces both BNs are present till the end of the pulse. In Fig. \ref{fig:7}(f) we also show the BN beating which reflects the switching-off of the 16.79~GHz BN. If other processes except rectification are causing the generation of this BN beating, it could in principle be used to lock both BNs to each other by a single, much lower frequency.

In the following, we concentrate on the down-mixed signal in Fig. \ref{fig:7}(d,e). The red shift of the 15.12~GHz BN can intuitively be understood by temperature-dependent group refractive index changes. The corresponding change in the group refractive index, which would lead to shifts of the order of 10 MHz, is estimated to be $\Delta n_g \approx \Delta f_{rep}/f_{rep}\cdot n_g = 0.0027 \Rightarrow \Delta n_r \approx 0.0006$. This corresponds to a reasonable 8~K change in temperature by assuming a temperature tuning coefficient $\beta_{tun}(K^{-1}) = \frac{1}{\lambda}\frac{d\lambda}{dT}\approx \frac{1}{n_{eff}}\frac{dn_{eff}}{dT} \approx 2\cdot 10^{-5} K^{-1} $ \cite{Ajili:EL:02:1675}.

Similar arguments for the BN at 16.79~GHz would mean a \textit{decrease} of temperature of 2.2~K, which is unphysical. Possible reasons for the blue shift could be temperature induced changes to the individual active regions, which are spatially separate (see inset of Fig. \ref{fig:1}(a)), and therefore could change the gain induced dispersion or the gain-clamping of each individual active region.

\begin{figure}[tb]
	\centering
	\includegraphics[width=0.8\linewidth]{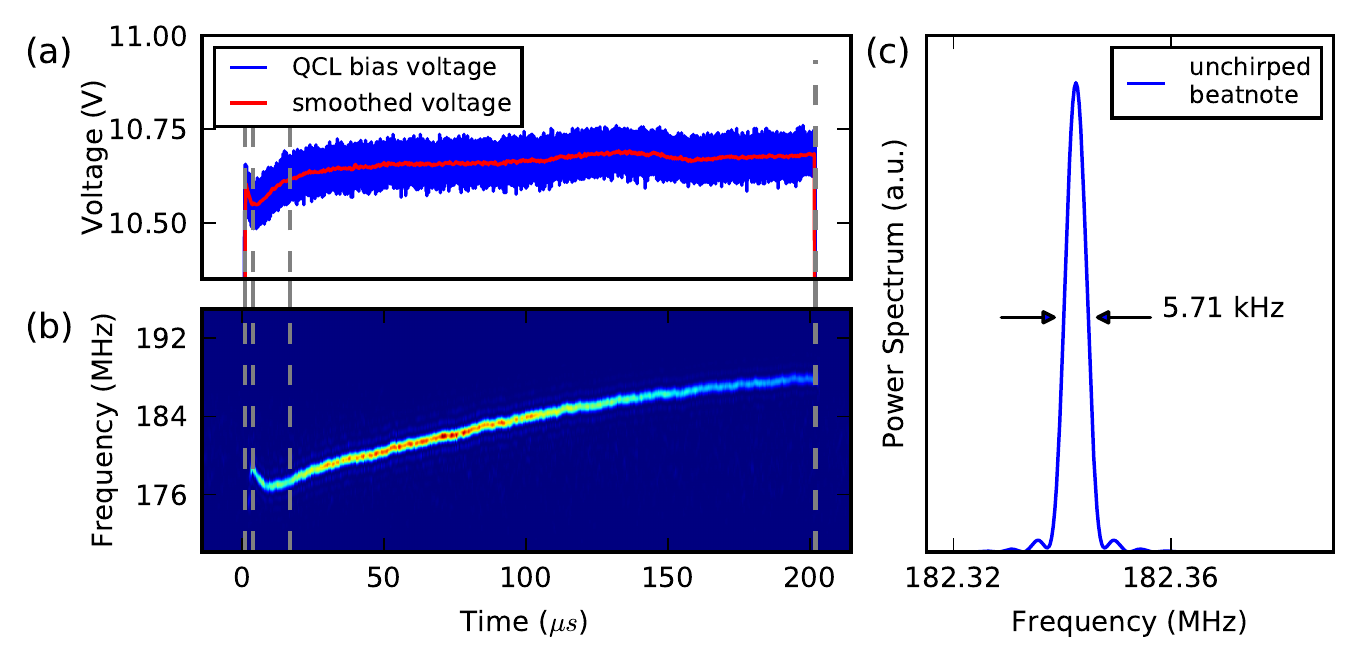}
	\caption{(a) Recorded QCL bias voltage over one 200~$\mu s$ long pulse. (b) Down-mixed 16.79~GHz BN of the same 200~$\mu s$ long pulse. Some trend from the voltage shows similarities to instantaneous frequency. (c) Numerically unchirped 16.78~GHz BN from the 2.3~THz FC which shows a narrow linewidth of 5.71~kHz.}
	\label{fig:8}
\end{figure}

So far, we assumed the current pulse to have a perfect rectangular shape. Since a bias-tee (containing a capacitor and inductor) is used for RF extraction from the bias line an additional component in the circuit is added. Fig. \ref{fig:8}(a) shows the QCL voltage time trace recorded with a fast oscilloscope and in (b) the corresponding down-mixed BN time trace in a single BN regime of the QCL. Slight changes in the voltage reflect or cause changes in the gain and would explain the common trend of the voltage and the BN drift. Whether the voltage affects the gain, and therefore the BN, or vice versa, is not yet known and the further investigation into this peculiar behavior is out of the context of this paper.

Due to these frequency drifts during the pulse the BN linewidths are artificially broadened. Now, by numerically compensating the drift, the BN linewidths can be estimated. The procedure uses the Hilbert-transformed electrical signal to get an analytical representation. The instantaneous frequency is derived from it and can be therefore compensated by multiplying this analytical signal by the opposite phase change referenced to a fixed frequency. Mathematically we perform the following steps:

\begin{align*}
	\text{fast scope signal:} \qquad & X(t) \xrightarrow{\text{Hilbert Transform}} X(t)+ 1i\cdot Y(t) \quad  & \text{: analytical signal}\\
	(\sim 180~MHz) \qquad & & &\\
	\Rightarrow & X(t)+1i\cdot Y(t)  = A(t)\cdot \exp [i\phi(t)]  & \text{: complex notation}\\[10pt]
	\Rightarrow & \nu(t) = \frac{1}{2 \pi} \cdot \frac{d\phi(t)}{dt} & \text{: instantaneous frequency}\\[10pt]
	\Rightarrow & \tilde{X}(t) = A(t)\cdot \exp [i\phi(t)] \cdot\\
	& \qquad \qquad \exp [i2\pi(\nu_{const}-\nu(t))t] & \text{: phase corrected scope signal}\\[10pt] 
\end{align*}

This approach is valid for the assumption of a single frequency signal as we have in our case. By Fourier Transforming the drift-free time signal we can extract narrow linewidths of 11~kHz for the BNs in Fig. \ref{fig:7}(a,b) and 14.9~kHz resp. 17.1~kHz for the 15.14~GHz resp. 16.79~GHz BN in Fig. \ref{fig:7}(d) resp. (e).
Applying the drift compensation to the 200~$\mu s$ pulse from Fig. \ref{fig:8}(a) even leads to linewidth estimates of 5.71~kHz, getting close to sub-kHz widths as presented in Fig. \ref{fig:8}(c).

\newpage

\section{Conclusion}

We show the simultaneous lasing of two octave-spaced FCs (at 2.3 and 4.6 THz) in a monolithic QCL operated in a $\mu s$ pulsed regime. By considering the cavity free spectral range of multiple devices at these two operating frequencies, taking into account the material dispersion of GaAs, we are able to predict and attribute the observed BNs frequencies to the 2.3 and 4.6 THz combs. This we then experimentally confirm for a 2.4~mm device by intermode beatnote spectroscopy, both conventionally using a Schottky diode, and also using a self-mixing technique. 
As we operate these lasers in pulsed operation, we are able to observe the $\mu s$ BN dynamics. With this technique, we show that the combs operate simultaneously, and by numerically compensating the drift, we estimate a BN linewidth of less than 15~kHz. Moreover, these traces reveal interestingly that, while one of the two BNs shows the conventional thermally-induced chirp, counterintuitively, the other chirps both up and down. We speculate this to be the result of a combination of factors, including spatial thermal effects acting differently on the gain for the 2.3 and 4.6 THz, the active regions for which are stacked one on top of the other. Furthermore, we observe what could be a weak coupling between the two combs, with the dynamics of one BN reflected on the other. 
By proving that this laser can output simultaneously a pair of combs spaced by an octave, one being close to the reststrahlen band even, we show this monolithic approach to be a promising platform for further development. With an active region tuned to lase CW and engineered for an optimized $\chi^{(2)}$, such an approach could lead to a fully integrated f-2f interferometer.

\section*{Funding}
H2020 European Research Council (724344); Schweizerischer Nationalfonds zur F\"orderung der Wissenschaftlichen Forschung (200020-165639).

\section*{Acknowledgments}
We would like to thank Urban Senica for the simulation of the waveguide group refractive index and David Stark for the discussion of the paper.


\bibliographystyle{osajnl}

\bibliography{bib_total,bib_total_AF, bib_total_AF3}

\end{document}